\documentclass[a4paper,prl,final,superscriptaddress,twocolumn]{revtex4}
\usepackage[latin1]{inputenc}
\usepackage[T1]{fontenc}
\usepackage[english]{babel}
\usepackage[draft]{graphicx}
\usepackage{amsmath}
\usepackage{amssymb}
\usepackage{mathrsfs}
\usepackage{textcomp}
\usepackage{tabularx}
\usepackage{color}
\usepackage{array}
 \usepackage{float}

\definecolor{darkblue}{rgb}{0.1,0.2,0.6}
\definecolor{darkred}{rgb}{0.8,0.1,0.2}
\usepackage[colorlinks,citecolor=darkblue,linkcolor=darkred,urlcolor=darkblue]{hyperref}

\begin{document}

\title{0-$\pi$ quantum transition in a carbon nanotube Josephson junction: universal phase dependence and orbital degeneracy.}
\author{R. Delagrange \footnote{raphaelle.delagrange@gmail.com}}
\affiliation{Laboratoire de Physique des Solides, CNRS, Univ. Paris-Sud, Université Paris Saclay, 91405 Orsay Cedex, France.}
\author{R. Weil}
\affiliation{Laboratoire de Physique des Solides, CNRS, Univ. Paris-Sud, Université Paris Saclay, 91405 Orsay Cedex, France.}
\author{A. Kasumov}
\affiliation{Laboratoire de Physique des Solides, CNRS, Univ. Paris-Sud, Université Paris Saclay, 91405 Orsay Cedex, France.}
\author{M. Ferrier}
\affiliation{Laboratoire de Physique des Solides, CNRS, Univ. Paris-Sud, Université Paris Saclay, 91405 Orsay Cedex, France.}
\affiliation{Department of Physics, Graduate School of Science, Osaka University, 1-1 Machikaneyama, Toyonaka, 560-0043 Osaka, Japan.}
\author{H. Bouchiat} 
\affiliation{Laboratoire de Physique des Solides, CNRS, Univ. Paris-Sud, Université Paris Saclay, 91405 Orsay Cedex, France.}
\author{R. Deblock \footnote{deblock@lps.u-psud.fr}}
\affiliation{Laboratoire de Physique des Solides, CNRS, Univ. Paris-Sud, Université Paris Saclay, 91405 Orsay Cedex, France.}

\begin{abstract}
We investigate experimentally the supercurrent in a clean carbon nanotube quantum dot, close to orbital degeneracy, connected to superconducting leads in a regime of strong competition between local electronic correlations and superconducting proximity effect. For an odd occupancy of the dot and intermediate coupling to the reservoir, the Kondo effect can develop in the normal state and screen the local magnetic moment of the dot. This leads to singlet-doublet transitions that strongly affect the Josephson effect in a single-level quantum dot: the sign of the supercurrent changes from positive to negative (0 to $\pi$-junction). In the regime of strongest competition between the Kondo effect and proximity effect, meaning that the Kondo temperature equals the superconducting gap, the magnetic state of the dot undergoes a first order quantum transition induced by the superconducting phase difference across the junction. This is revealed experimentally by anharmonic current-phase relations. In addition, the very specific electronic configuration of clean carbon nanotubes, with two nearly orbitally degenerated states, leads to different physics  depending whether only one or both quasi-degenerate upper levels of the dots participate to transport, which is determined by their occupancy and relative widths.
When the transport of Cooper pairs takes place through only one of these levels, we find that the phase diagram of the phase-dependent 0-$\pi$ transition is a universal characteristic of a discontinuous level-crossing quantum transition at zero temperature. In the case were two levels participate to transport, the nanotube Josephson current exhibits a continuous 0-$\pi$ transition, independent of the superconducting phase, revealing a different physical mechanism of the transition.\newline

Subject areas: Condensed matter physics: mesoscopics and superconductivity, Quantum physics

\end{abstract}

\maketitle

\section{Introduction}

Understanding and controlling the superconducting proximity effect in various systems, from ferromagnetic metals to topological insulators, is a subject of great interest. Indeed, Josephson junctions - referring to any non superconducting material sandwiched between two superconductors - have become the main ingredient of superconducting circuits and quantum electronics. 
The simplest Josephson junction (JJ), an insulator between two superconductors, is passed through by a non dissipative current $I=I_C \sin(\varphi)$, $I_C$ being the critical current and $\varphi$ the phase difference of the superconducting order parameters. When a normal metal is inserted between the superconductors, the transmission of Cooper pairs takes place through Andreev Bound States (ABS), that are confined in the normal region at an energy below the gap \cite{NazarovBlanter}. Due to the boundary conditions, the energy of these bound states depends on $\varphi$, leading to a phase dependence of the supercurrent: the current-phase relation (CPR). Since the supercurrent is strongly affected by the physics of the normal part, the CPR is a powerful probe of the interactions and correlations in the system.  In this work, we investigate quantum dot (QD) Josephson junctions, where Coulomb interactions cause the CPR to be highly dependent on the dot's occupancy and on the number of energy levels involved in the transport.

In a single level QD-JJ, the physics is governed by four characteristic energies: the coupling $\Gamma=\Gamma_L+\Gamma_R$ ($\Gamma_L$ and $\Gamma_R$ are the coupling respectively to the left and right reservoirs, $\Gamma_L/\Gamma_R$ is the asymmetry), the charging energy $U$, the level energy in the dot $\epsilon$ and the superconducting gap of the contacts $\Delta$. We focus in this article on the intermediate regime $\Gamma\approx U\approx\Delta$, where the Coulomb Blockade is strong enough to impose a well defined occupancy and the coupling sufficient to observe a supercurrent \cite{defranceschi2010}.  The transfer of Cooper pairs then involves cotunneling processes, strongly dependent on the dot's occupancy. When this occupancy is even, one has a 0-junction whose amplitude follows the transmission of the dot. But for an odd occupancy, the first non-zero contribution to the supercurrent involves forth order processes, that imply reversing the spin-ordering of the Cooper pair. The sign of the supercurrent is reversed and its amplitude strongly reduced, this is called a $\pi$-junction. Experimentally, the supercurrent can be precisely changed tuning the parity of the dot with a gate voltage \cite{Vandam2006,Cleuziou2006,Jorgensen2007}. 
 
 In addition, an oddly occupied dot gives rise to Kondo effect. The interaction of the local magnetic moment with delocalized conduction electrons through spin flip processes leads, in the normal state, to the formation of a strongly correlated state. This Kondo singlet state is characterized by the screening of the dot's magnetic moment and by a resonance in the density of states at the Fermi energy for temperatures below the Kondo temperature $T_K$ \cite{Pustilnik2004,Goldhaber-Gordon1998,Cronenwett1998}. In the superconducting state, when $k_B T_K < \Delta$, the Kondo screening is destroyed by superconducting correlations and does not affect the $\pi$-junction. But for $k_B T_K\gg\Delta$, the unpaired electron's spin is involved in a Kondo singlet that opens a well transmitted channel in the system and facilitates the transfer of Cooper pairs. Therefore the 0-junction is recovered and the supercurrent is enhanced due to the cooperation between superconductivity and Kondo effect \cite{Clerk2000,Glazman1989,Jorgensen2007,Eichler2009}. Since the Kondo temperature depends on $U$, $\Gamma$ and $\epsilon$ \cite{Haldane1978}, the junction can be tuned from 0 to $\pi$ by varying these parameters for a fixed parity and value of $\Delta$ \cite{Maurand2012}.
Measuring the CPR of the single-level QD Josephson junction directly gives insights into the magnetic state of the system: a doublet if one measures a $\pi$-junction, a singlet state (purely BCS or Kondo) otherwise. Here, we are particularly interested in the specific regime of the 0-$\pi$ transition, where the system undergoes a level-crossing quantum transition. The fundamental state of the system, 0 or $\pi$, depends on the superconducting phase $\varphi$, meaning that the magnetic state of the system (singlet or doublet) can be controlled by this parameter \cite{Vecino2003,choi2004,Siano2004,Bauer2007,Karrasch2008,Meng2009,Luitz2010,Luitz2012,Delagrange2015}.
 \newline

In a multi-level quantum dot, this simple picture is not valid anymore, as shown predicted theoretically \cite{Rozhkov2001,Zazunov2010,Lee2010,Karrasch2011,Droste2012}. The measurement of the current-phase relation is no longer a good indicator of the effective magnetic moment of the dot. Indeed, as soon as several energy levels participate to the transport, the available cotunelling processes are different and the properties of the wave-functions become determinant, making 0 and $\pi$-junction possible both for even and odd occupations. This multi-level effects on the supercurrent in a quantum dot based Josephson junction have been experimentally observed by van Dam et al. \cite{Vandam2006} with an InAs nanowire in which, unlike in carbon nanotube, the exact electronic configuration is not known. \newline

The aim of this article is to investigate the CPR in a clean carbon nanotube (CNT) QD, where the orbital levels are nearly degenerated.
Whereas a number of specific effects of this orbital degeneracy have been pointed out in the normal state \cite{Laird2014}, their signature on the Josephson effect in CNT has not been explored yet.
  
Our results show that distinct behaviors emerge depending on the number of levels involved in transport. This number is determined by the occupancy and the relative widths of the nearly degenerated orbital levels. In our sample, for most filling factors, the system is well understood in a single-level description. In this regime, for odd electronic occupation and intermediate transmission of the contacts, we have experimentally proven the phase-controlled 0-$\pi$ transition \cite{Delagrange2015} and compared it with Quantum Monte Carlo calculations. We show in this article that this first-order quantum transition happens for a characteristic superconducting phase that has a universal behavior. On the other hand, in some odd diamonds with nearly degenerated orbital levels, we qualitatively confirm theoretical predictions about the gate dependence of the supercurrent in the two-level regime and its high sensitivity to the precise configuration of the two orbital states involved in transport. In addition, the phase dependence of the supercurrent shows a continuous 0-$\pi$ transition with a complete cancellation of the amplitude of the Josephson current, in contrast with the first order single-level 0-$\pi$ transition.

\section{Experimental setup and characterization in the normal state}

\subsection{Experimental setup: a carbon nanotube inserted in a SQUID}

 \begin{figure}[h]
    \begin{center}
    \includegraphics[width=6cm]{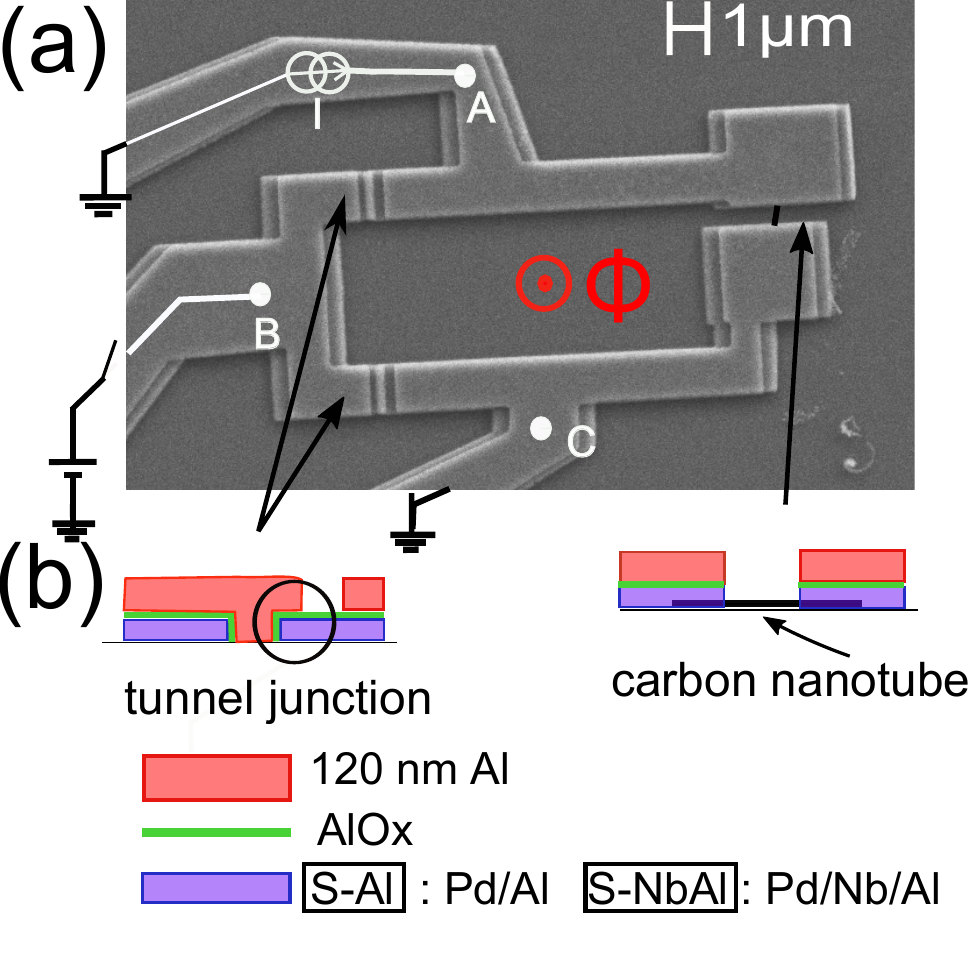}
    \end{center}
    \caption{Color. (a) Scanning electron microscopy image of the measured asymmetric SQUID, containing two reference JJs in parallel with a CNT based Josephson junction (see text and \cite{Basset2014}). To phase bias the CNT junction, a magnetic flux $\Phi$ is applied with to a magnetic field perpendicular to the SQUID. (b) Schematics of the layers constituting the tunnel junctions and the contacts of the CNT. The first layer of sample S-Al is made of Pd(7 nm)/Al(70). The first one of S-NbAl is made of Pd(7~nm)/Nb(20~nm)/Al(40~nm).}
    \label{sample}
    \end{figure}
The sample is made of a CNT quantum dot contacted with superconducting leads, such that it can be passed through by a supercurrent \cite{Kasumov1999}. Compared to semi-conducting nanowires or two-dimensional electron gases, clean carbon nanotubes benefit of well transmitted superconducting contacts and a (quasi) orbital degeneracy. They thus realize model systems for the investigation of two-level physics in QDs.
 
  \begin{figure*}[htpb]
     \begin{center}
     \includegraphics[width=17cm]{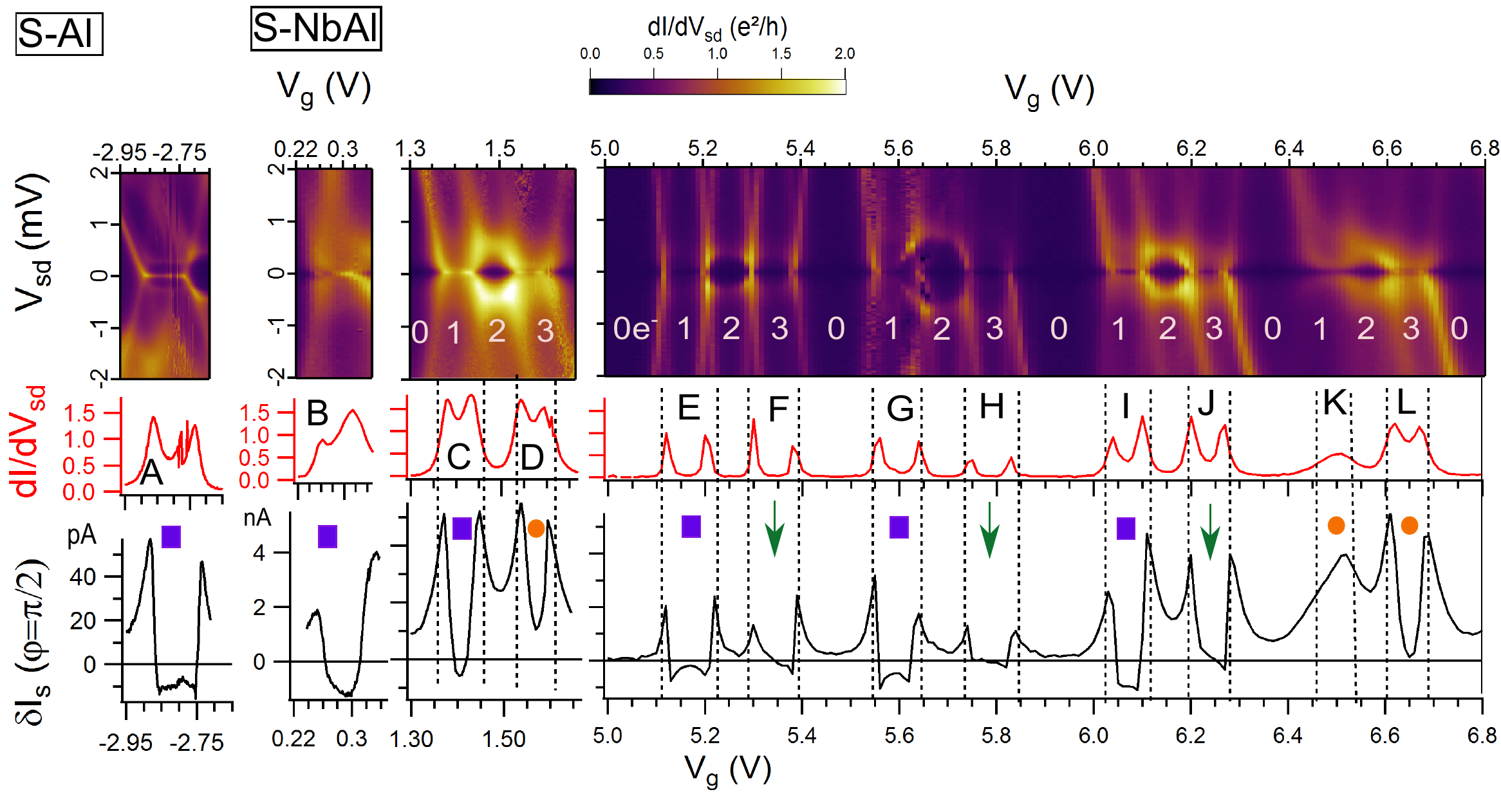}
     \end{center}
     \caption{Two columns. Color. Colorplot: Differential conductance $dI/dV_{sd}$ as a function of $V_{sd}$ and $V_g$ for both samples S-Al and S-NbAl in the normal state, applying $B=0.13\mathrm{~T}$ for S-Al (diamond A) and $B=1\mathrm{~T}$ for S-NbAl (diamonds B to L). In white is written the number of electrons in the last occupied shell. 
     Red graphs: horizontal cuts showing $dI/dV_{sd}$ at zero bias versus gate voltage $V_g$ in the normal state. 
     Black graphs: supercurrent $\delta I_s(V_g)$ for a superconducting phase difference of $\varphi=\pi/2$. Orange circles indicate Kondo induced 0-junction, for odd occupancies. Purple squares show $\pi$-junctions with a single-level behavior. The green arrows indicate two-level 0/$\pi$-junctions. The consequences of two-level physics are more spectacular for F and J than for H, where $\delta E$ is larger than in F and J (see table I).}
     \label{dIdV_Is}
     \end{figure*} 
 
 This CNT-based QD is embedded in an asymmetric modified SQUID (Fig. \ref{sample}). In one branch of the SQUID is the QD JJ (here the CNT). The other branch contains a reference JJ with a large critical current compared to the one of the QD JJ, allowing to determine the CPR of interest \cite{DellaRocca2007,Basset2014}. The switching current $I_s$ of the SQUID is measured as a function of the magnetic flux through the SQUID, which is proportional to the phase $\varphi$ across the QD JJ. Its CPR is then obtained by extracting the modulation of $I_s$ around its mean value $\left<I_s \right>$, called $\delta I_s$. Our device possesses a second reference JJ and a third connection as described in Ref. \cite{Basset2014,Delagrange2015}. This enables to independently characterize each junction at room temperature and to measure both the CPR of the CNT and its differential conductance in the superconducting state [Appendix III].
 
The CNTs are grown by chemical vapor deposition on an oxidized doped silicon wafer \cite{Kasumov2007}. The contacts of the nanotube are separated by a distance $L=400\mathrm{~nm}$ and are made of aluminum-based multi-layers. Two samples are presented in this article: the first one, called S-Al, is contacted with a Pd(7~nm)/Al(70~nm) bilayer, whose superconducting gap is $\Delta_{PdAl} = 65 \pm 5 \mathrm{~µeV}$. This value is considerably reduced compared to pure aluminum $\Delta_{Al} \approx 200\mathrm{~µeV}$ due to the palladium layer, that provides good contacts on the CNT. For the second sample, called S-NbAl, we added a niobium layer between palladium and aluminum, in order to increase the gap to $\Delta_{PdNbAl} = 170 \pm 5 {~µeV}$: Pd(7~nm)/Nb(20~nm)/Al(40~nm) (see Fig.\ref{sample}). The two Josephson junctions in the other branch of the SQUID are made by oxidation of the first aluminum layer followed by angle deposition of a second Al layer (120~nm).

 The sample is then cooled down in a dilution refrigerator of base temperature 50 mK and measured through low-pass filtered lines. The phase difference across the CNT-junction $\varphi$ is controlled applying a magnetic field $B$ perpendicularly to the sample. $\varphi$ is then proportional to $B$ with $\varphi=2\pi B S/ \Phi_0$, $\Phi_0=h/2e$ being the superconducting flux quantum and $S\approx 40\mathrm{~\mu m}^2$ the loop area. 
 
\subsection{Characterization in the normal state}
A CNT QD can sustain different regimes, depending on the values of the coupling $\Gamma$ and the charging energy $U$: as $\Gamma$ increases compared to $U$, the transport regime goes from pure Coulomb blockade to Kondo effect, and finally Fabry-Pérot regime \cite{Makarovski2007}. To characterize the system, we measured the differential conductance $dI/dV_{sd}$ as a function of the bias voltage ($V_{sd}$) and the gate voltage ($V_g$) in the normal state. This is done using a lock-in-amplifier technique (with a modulation of $20\mathrm{~\mu V}$) and applying a magnetic field large enough to destroy the superconductivity in the contacts. For S-Al, contacted with Pd/Al, a magnetic field $B=0.13\mathrm{~T}$ is enough while, for S-NbAl, we needed to apply $B=1\mathrm{~T}$ due to the niobium layer, whose critical field is higher. The resulting stability diagrams of both samples are represented on Fig.\ref{dIdV_Is}.
The sample S-Al exhibits Coulomb diamonds. For the diamond shown (diamond A), there is a maximum of conductance at zero bias, indicating Kondo effect with a regular temperature dependence of the resonance (see Appendix I). Two satellite peaks at finite $V_{sd}$ indicate inelastic cotunneling.

 \begin{figure}[h]
    \begin{center}
    \includegraphics[width=6cm]{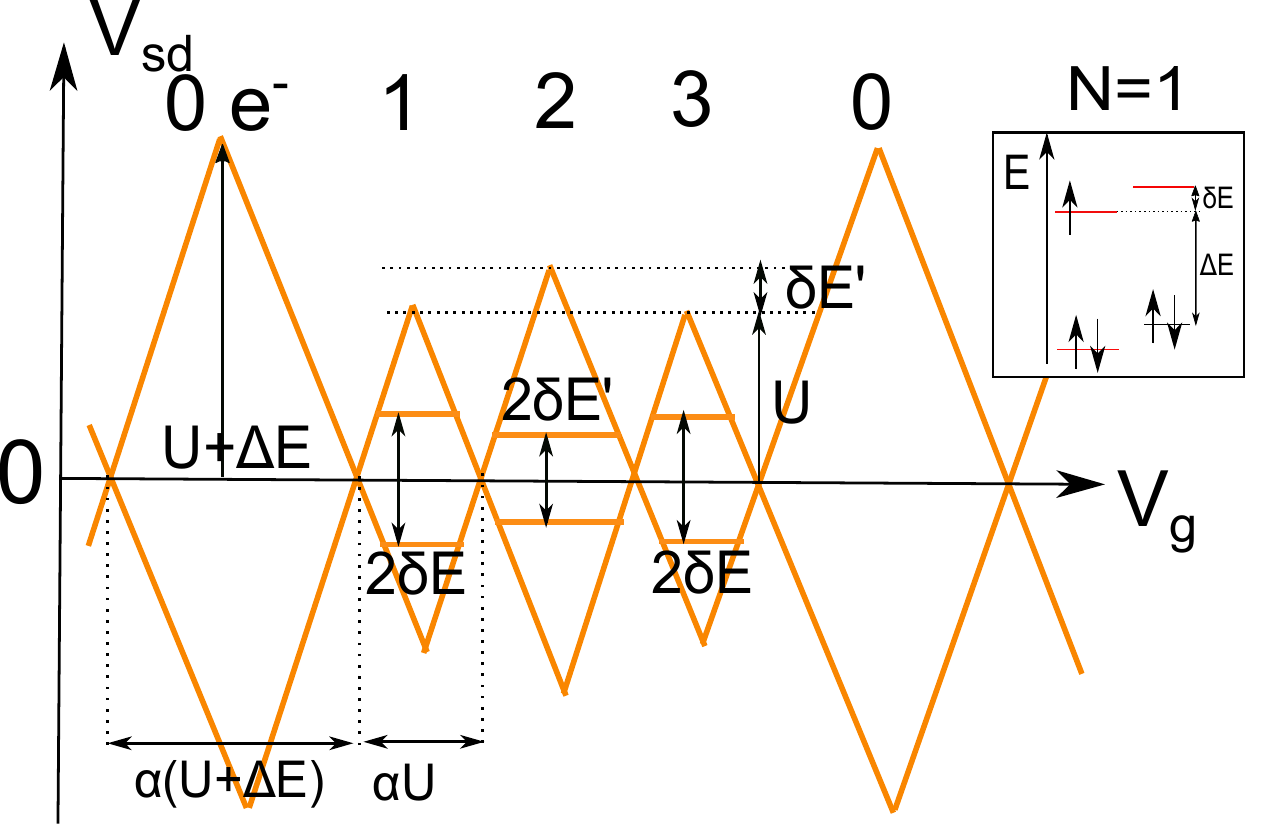}
    \end{center}
    \caption{Color. Typical stability diagram expected in a clean CNT, of charging energy U, where the energy levels are separated by $\Delta E$ and the orbital degeneracy is lifted by $\delta E$. The occupancy is indicated on the top of each Coulomb diamond. We call $\alpha$ the proportionality factor between the energy level $\epsilon$ and the applied gate voltage $V_g$. The spacing between the inelastic cotunneling peaks in the double occupancy (N=2) is called $2\delta E'$. $\delta E'\neq\delta E$ because of exchange interactions \cite{Babic2004}. Inset: corresponding electronic configuration for N=1.}
    \label{diamond_CNT}
    \end{figure}
    
In the sample S-NbAl, Coulomb diamonds are present on a large range of gate voltage (diamonds B to L and more, not shown) with a four-fold degeneracy \footnote{Note that, on the figure, due to the small range of $V_{sd}$, one cannot see complete closure of the diamonds, we nevertheless use this term to designate a range of gate voltage between two Coulomb peaks.}. These features are typical of a clean CNT, where each energy level is nearly orbitally degenerated, in addition to the spin degeneracy. This enables the determination of the electronic occupation of the highest occupied shells: N=0, 1, 2 or 3 electrons (white numbers on Fig. \ref{dIdV_Is}). The presence of cotunneling peaks at finite energy indicates that the orbital degeneracy is lifted, either due to spin-orbit interactions, boundary conditions or defects in the nanotube \cite{Kuemmeth2008,Jespersen2011,Marganska2015}. The relevant energies of the system can be determined from this stability diagram (Fig. \ref{diamond_CNT} and Appendix I): $\Delta E$ the spacing between two successive energy levels in the CNT, $U$ the charging energy and $\delta E<\Delta E, U$ the lift of degeneracy \cite{Babic2004,Holm2008}.
In most oddly-occupied diamonds of S-NbAl, the conductance is non-zero at zero-bias, suggesting the presence of Kondo effect (diamonds B-D and I-L). The expected Kondo resonance is not as clear as in the diamond A, because of the 1T magnetic field needed to destroy the superconductivity in the Pd/Nb/Al contacts, that affects the Kondo effect (see Appendix I). The determination of $T_K$ or $\Gamma$ is thus more tricky, their evaluation is explained in Appendix I.
The values of the dot's parameters are given in Table I, for all the oddly occupied diamonds investigated.

\begin{table*}
\renewcommand\arraystretch{1.8}
\begin{center}
\begin{tabular}{|c||c|c|c|c|c|c|c|c|c|c|c|c|}
\hline
~  & A & B & C & D & E & F & G & H & I & J & K & L \\
\hline
\hline
Occupancy (N=) & 1 & 1 & 1 & 3 & 1 & 3 & 1 & 3 & 1 & 3 & 1 & 3 \\
\hline
U (meV) & 1.6 & 2.8 & 2.3 & 2.3 & 3.9 & 3.9 & 3.4 & 3.4 & 3.2 & 3.2 & 2.3 & 2.3\\
\hline
$\Gamma$ (meV) & 0.25 & 0.43 & 0.5 & 0.5* & \multicolumn{2}{c|}{0.35*} & \multicolumn{2}{c|}{0.4*} & 0.44 & 0.45* & \multicolumn{2}{c|}{0.55*} \\
\hline
$\Delta E$ (meV) & 1 & 5 & \multicolumn{2}{c|}{4} & \multicolumn{2}{c|}{4} & \multicolumn{2}{c|}{5} & \multicolumn{2}{c|}{4} & \multicolumn{2}{c|}{3.5} \\
\hline
$\delta E$ (meV) & 0.3 & 0.8 & 0.5 & 0.2 & 0.4 & 0.4 & 0.7 & 0.7 & 0.3 & 0.3 & 0.25 & 0.25 \\
\hline
$T_K$ (meV) & 0.036 & 0.06 & 0.13 & 0.13* & \multicolumn{2}{c|}{0.01*} & \multicolumn{2}{c|}{0.03*} & 0.048 & 0.05* & \multicolumn{2}{c|}{0.15*} \\
\hline
$\Delta$ (meV) & 0.064 (Pd/Al) &  \multicolumn{11}{c|}{0.17 (Pd/Nb/Al)} \\
\hline

\end{tabular}
\end{center}
    \label{tableau}
    \caption{Two columns. Quantum dot's parameters for different gate voltage regions: the charging energy U, the coupling $\Gamma$, $\Delta E$ the level-spacing, $\delta E$ the lift of orbital degeneracy, $T_K$ the Kondo temperature calculated with formula \ref{Tk} (at $\epsilon=0$) and the superconducting gap $\Delta$. The values with a * are known only with a 20\% uncertainty, while the other values are given within a 10\% uncertainty.}
\end{table*}

For S-NbAl, $\Delta E$ is typically equal to 4-5 meV, a value reasonably consistent with the expression $\Delta E\approx \frac{hv_F}{2L}$, with $v_F\approx 8.10^5~\mathrm{m/s}$ the Fermi velocity \cite{Charlier2007} for a metallic nanotube and $L=400\mathrm{~nm}$. The value of $\delta E$ varies between $0.2$ and $0.8\mathrm{~meV}$, which is small compared to U and of the order of $\Gamma$, suggesting that two-level physics is relevant \cite{Vandam2006}. Note that, for the range of gate voltage presented, there is no signature of SU(4) Kondo effect, that would be characterized by a strong Kondo effect spreading over N=1, N=2 and N=3 diamonds and a conductance that can reach $4e^2/h$ \cite{Makarovski2007}. To observe this kind of exotic Kondo effect, the Kondo temperature should be very large compared to the lift of degeneracy $\delta E$ \cite{Cleuziou2013}. In Appendix IV, we present measurements in a range of gate voltage where this condition is to be fulfilled.

\section{Superconducting state : Gate controlled 0 or $\pi$-junction and two-level physics.}

By switching off the magnetic field, we measure the current-phase relation (CPR) of the QD JJ in the superconducting state. 
For this, the SQUID is biased with a linearly increasing current\footnote{Rate of the current ramp: $\frac{\mathrm{d}I}{\mathrm{d}t}=37~\mathrm{\mu A/s}$}, the switching current is directly obtained from the time at which the SQUID switches to a dissipative state. This process is reproduced and averaged around 1000 times, the whole procedure being repeated at different values of magnetic field below a few Gauss. This experiment probes the switching current of the SQUID, which is always smaller than the critical current $I_c$. To obtain its modulation $\delta I_s$ versus the magnetic field, the constant contribution of the reference junctions is subtracted.
As demonstrated in Ref. \cite{Basset2014}, $\delta I_s(B)$ is proportional to the CPR of the CNT junction, with a proportionality factor lower than one. One should note that this detection scheme, in particular near the 0-$\pi$ transition, is very sensitive to the electromagnetic environment, which therefore needs to be optimized (see Appendix II).

We first focus on the measurement of the supercurrent at a fixed phase $\varphi =\pi /2$, represented as a function of the gate voltage on Fig. \ref{dIdV_Is} (black curves). This quantity is proportional to the critical current of the junction in the case of sinusoidal CPRs (which represent most of the CPRs at finite temperature), with an additional information on the sign.

For the diamonds with an even electronic occupancy, this supercurrent is always positive, indicating 0-junctions. In contrast, for odd occupancies, there are two situations: for some diamonds, A, B, E, F, G, H, I and J (blue squares and green arrows on Fig. \ref{dIdV_Is}), the supercurrent is negative around the center of the diamond and has a reduced amplitude:  this is the signature of a $\pi$-junction. But, for the others odd diamonds (D, K, L, orange circles on Fig. \ref{dIdV_Is}, and all the diamonds for $V_g \geq 6.8\mathrm{ V}$ (not shown), the supercurrent does not change sign. These odd parity 0 junctions are attributed to the Kondo effect that screens the magnetic moment of unpaired electrons. This hypothesis is corroborated by the comparison of the ratio $\frac{\Delta}{T_K}$ for the different diamonds (see table I): when $T_K\ll \Delta$, the Kondo correlations are destroyed by the superconducting proximity effect, the junction turns $\pi$. But when $T_K\geq \Delta$, as for diamonds C, D, K, L and $V_g \geq 6.8\mathrm{ V}$, the supercurrent is enhanced by the cooperation of the Kondo effect and superconductivity, leading to a 0-junction. 

However, some of the odd parity $\pi$-junctions (F, H and J, green arrows on Fig. \ref{dIdV_Is}), of filling factors N=3, reveal unusual features. For typical single-level $\pi$-junctions as the ones indicated by blue squares, the supercurrent is symmetric relatively to the center of the diamond \cite{Vandam2006,Luitz2012}, which corresponds to the half-filling point in a single-level description. But for the diamonds indicated by green arrows, the supercurrent $\delta I_s$ has a completely different behavior. All over the left side of the diamond (close to the N=3 to 2 degeneracy point), the supercurrent is positive. This 0-junction in an odd-occupied diamond cannot be attributed to the Kondo effect, that is not strong enough to turn the junction from $\pi$ to 0, as in diamonds D and L \footnote{If the Kondo effect was strong enough for that, like in diamonds D and L, there would not be any occurrence of $\pi$-junction either in diamonds I or J.}. Therefore, it can only be attributed to the participation of a second energy level to the transport \cite{Rozhkov2001,Lee2010}. Close to the N=3 to 0 degeneracy point (right side of the diamond), the junction is in a $\pi$ state. The 0 to $\pi$ transition occurs around the center of the N=3 diamond, and the $\pi$-0 transition between the N=3 and N=4 diamonds.

 The supercurrent is asymmetric with respect to the center of the diamond. \footnote{The effect of two-level physics is more spectacular for F and J than for H, in which the two-level description is less relevant since $\delta E$ is almost two times larger than in F and J (see table I).} This symmetry breaking was predicted, but not yet measured, in the particular case of two-level quantum dots \cite{Shimizu1998} and carbon nanotubes \cite{Yu2010}. Indeed, in a two-level description of a quantum dot, the half-filling point is moved to the center of the N=2 diamond.
 The striking point of our experiment is that these peculiar 0-$\pi$ transitions, \textit{i.e} two-level behaviors, are observed (for three diamonds) at N=3 occupancies instead of both N=1 and N=3 as predicted by Refs. \cite{Shimizu1998,Yu2010}. This point will be addressed in the last part of this article. 
 
 Our experiment thus demonstrates the existence of two kinds of gate induced 0-$\pi$ transitions: some involves only single-level physics whereas others have signature of two-level physics. In the next two parts, we detail the phase dependence of the supercurrent in each regime. 

\section{Universality of the single-level phase induced 0-$\pi$ transition}

 In this part, we discuss the phase dependence of the supercurrent in the single-level regime.
The N=1 diamonds, which behave as single-level systems, are analyzed and interpreted. We show that the comparison of the various 0-$\pi$ transitions measured in this regime leads to an universal phase diagram of the transition.

\subsection{Detailed analysis of the single-level 0-$\pi$ transition}

 \begin{figure*}
          \begin{center}
          \includegraphics[width=17cm]{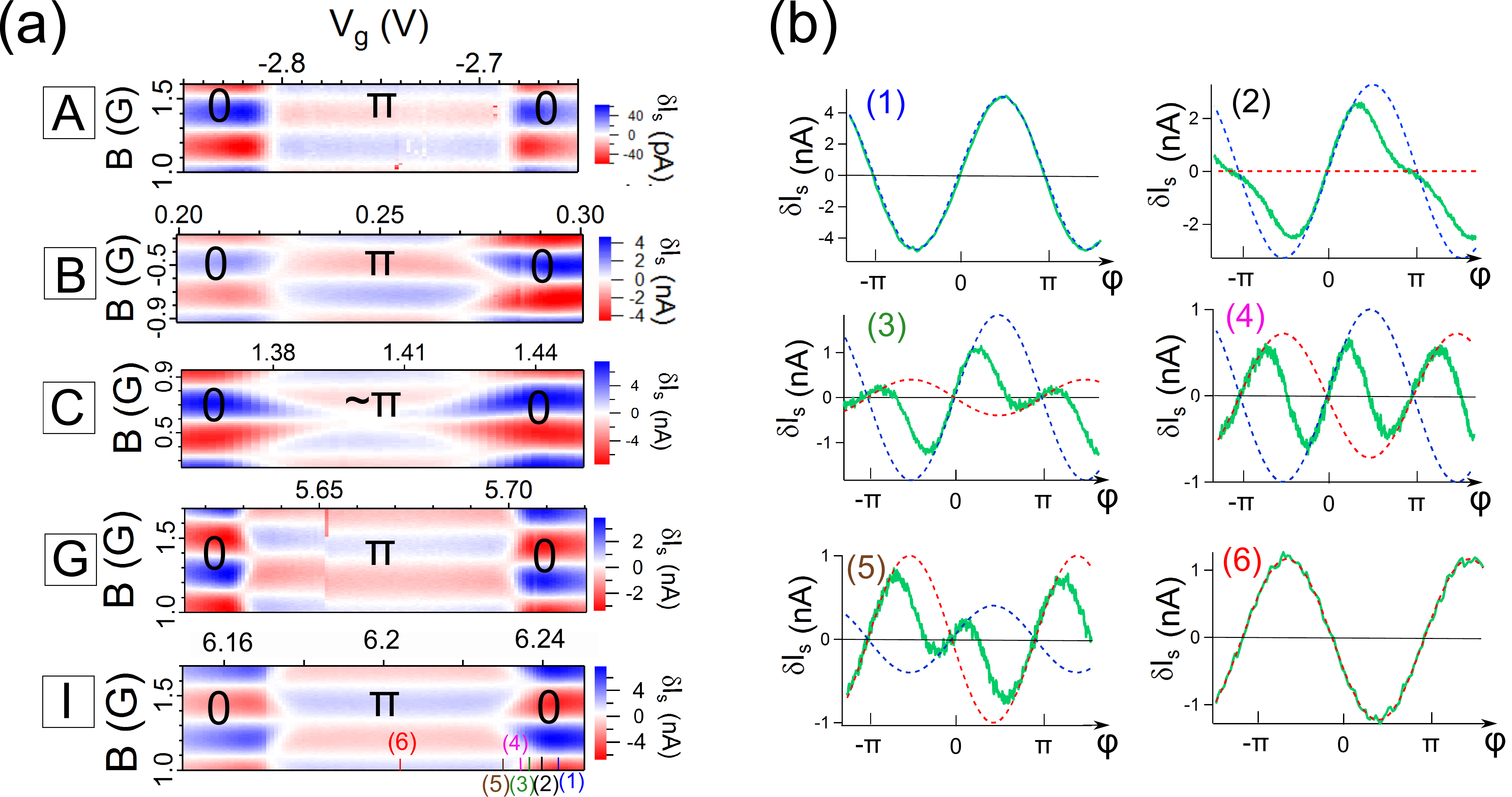}
          \end{center}
          \caption{Two columns. Color.(a): modulation of the switching current of the SQUID $\delta I_s$, proportional to the CPR,  as a function of the magnetic field B and the gate voltage $V_g$, for the diamonds A, B, C, G and I. Vertical cuts at the 0-$\pi$ transition are represented for ridge I, showing the whole transition. For the diamond G, the color-plot exhibits a discontinuity, probably due to the trapping of a vortex. (b): CPR, for diamond I, near the transition (green continuous line) versus the superconducting phase $\varphi$. The dashed lines are guides to the eyes and represent the contributions of the singlet (0-junction, in blue) and the doublet state ($\pi$-junction, in red).}
          \label{detail_cpr}
          \end{figure*}

In this section, we focus on oddly occupied diamonds in the single-level regime, that present in the superconducting state a $\pi$-junction centered in the middle of the diamonds and 0-junction on the edges: diamonds A, B, C, G and I. This is compatible with 0-$\pi$ transitions  $T_K\approx \Delta$, driven by the interplay between the Kondo effect and the superconductivity when, since $T_K$ is symmetric with respect to the center of the diamond, where it is minimal, and increases on the edges (formula \ref{Tk}).

The modulation of the switching current $\delta I_s$ versus magnetic field B, proportional to the CPR, is measured for various $V_g$ and is represented on Fig.\ref{detail_cpr} for some representative oddly-occupied ridges. Diamond A, from sample S-Al, presents a small supercurrent ($\approx 40 \mathrm{~pA}$) and a transition from 0 to $\pi$ that extends on a very small range of $V_g$ (smaller than $100\mathrm{~\mu V}$), beyond the precision of our experiment. In contrast, in sample S-NbAl, due to a larger superconducting gap, the $\pi$-junction supercurrent is larger ($\approx 4 \mathrm{~nA}$). Moreover, the width in gate voltage of the transition is larger, allowing to measure the CPR very accurately in the transition region. Noteworthy, the region C exhibits an incomplete 0-$\pi$ transition where the CPR is not completely reversed in the center of the diamond.

     \begin{figure}
             \begin{center}
             \includegraphics[width=\columnwidth]{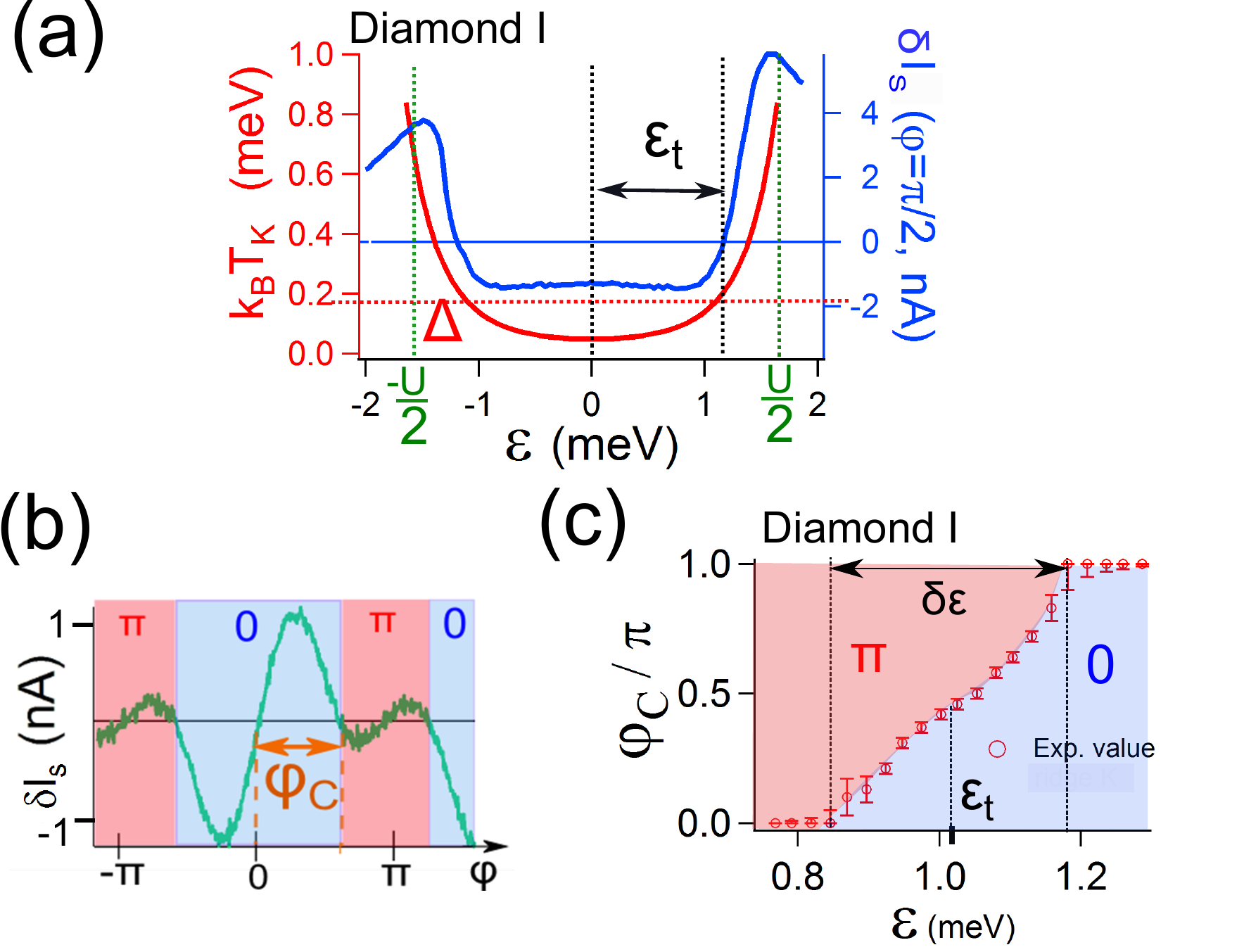}
             \end{center}
             \caption{Color. Phase dependent transition: definition of the relevant quantities (a) $T_K$ (red line), $\Delta$ (red dotted line) and $\delta I_S(\varphi= \pi/2)$ (blue line) for the diamond I as a function of the energy level $\epsilon$, defined as equal to zero at the middle of the diamond and such that, in a diamond, $\epsilon \in [-U/2,U/2]$. One can notice that the sign of $\delta I_s$ changes when $T_K\approx \Delta$. $\epsilon_t$ is the value of $\epsilon$ for which the junction switches from $\pi$ to 0 at $\varphi=\pi/2$.
             (b) Definition of the critical phase $\varphi_C$ such that the CPR has 0-behavior for $\varphi \in [0,\varphi_C]$ and $\pi$-behavior for $\varphi \in [\varphi_C,\pi]$. 
             (c) This quantity is plotted as a function of $\epsilon$ for diamond I, yielding a phase diagram of the $\varphi$-controlled transition. We call $\delta \epsilon$ the width of the transition.}
             \label{phi_c}
             \end{figure}
   
     \begin{table}[h]
         \renewcommand\arraystretch{1.8}
         \begin{center}
         \begin{tabular}{|c||c|c|c|c|c|c|c|c|c|}
         \hline
         meV & \multicolumn{2}{c|}{B} & \multicolumn{2}{c|}{C} & \multicolumn{2}{c|}{G} & \multicolumn{2}{c|}{I} & J right\\
         \hline
         \hline
         U  &  \multicolumn{2}{c|}{2.8} &  \multicolumn{2}{c|}{2.3} &  \multicolumn{2}{c|}{3.5} &  \multicolumn{2}{c|}{3.2}&  3.2 \\
         \hline
         $k_B T_K$& \multicolumn{2}{c|}{$\approx$ 0.06} &  \multicolumn{2}{c|}{0.13} &  \multicolumn{2}{c|}{$\approx$ 0.03} &  \multicolumn{2}{c|}{0.05} & 0.05 \\
         \hline
         $2\delta \epsilon/U$& 0.23 & 0.43 & 0.87 & 0.96 & 0.06 & 0.06 & 0.15 & 0.2 & 0.2 \\
         \hline
          $2\epsilon_t/U$& 0.79 & 0.64 & 0 & 0 & 0.74 & 0.74 & 0.75 & 0.69 & 0.8 \\
          \hline
         
         \end{tabular}
         \end{center}
             \caption{Values of U, $T_K$, $\delta \epsilon$ and $\epsilon_t$, given in meV for the investigated diamonds. For diamonds B, C, G and I, there are two transitions (0 to $\pi$ and $\pi$ to 0), with different parameters. In diamond J, only the right side of the diamond exhibits a phase dependence of the transition.}
              \label{table2}
         \end{table}

On Fig.\ref{detail_cpr} (b) are shown CPRs extracted from the 0-$\pi$ transition of diamond I.   
On the edges of the ridge, far from the transition (Fig. \ref{detail_cpr}.(b).(1)), the junction  behaves as a regular JJ 0-junction, with a CPR proportional to $\sin (\varphi)$. In contrast, at the center of the diamond (Fig. \ref{detail_cpr}.(b).(6)), where $T_K$ is minimum, the CPR is $\pi$-shifted ($\delta I_s \propto \sin(\varphi + \pi)$) and has a smaller amplitude, characteristic of a $\pi$-junction. In between, the CPR is anharmonic: a distortion appears first around $\pi$ and develops as $T_K$ decreases. The CPR is composite, with a part of type "0" around $\varphi=0$ and a $\pi$-junction behavior around $\varphi=\pi$. The transition from one part to the other is achieved varying the superconducting phase. The rounding of this transition by finite temperature was taken into account in QMC calculations, giving an excellent agreement for an electronic temperature of $150\mathrm{~mK}$ \cite{Delagrange2015}. These data are thus consistent with a phase controlled level-crossing quantum transition in a single-level QD \cite{Delagrange2015,Zonda2015}. In other words, one can control the magnetic state of the junction, doublet or singlet, with the superconducting phase.

 \subsection{Universal scaling of the critical phase}

 We present here a quantitative study of the level-crossing quantum transition in the single-level regime. We call $\varphi_C$ the superconducting phase at which, at a fixed gate voltage, the system undergoes the transition from 0 to $\pi$. Theoretically, this critical phase $\varphi_C$ is defined at T=0, where the transition is expected as a jump in the supercurrent. At finite temperature the transition is rounded but, if  T is small enough, $\varphi_C$ equals the phase at which the supercurrent is zero \cite{Karrasch2008,Delagrange2015} (see Fig. \ref{phi_c} (b)). On Fig. \ref{phi_c}.(c), this quantity is represented for the diamond I as a function of the level energy $\epsilon$, proportional to $V_g$, giving a phase diagram of the phase mediated transition.  
 
 \begin{figure}
              \begin{center}
              \includegraphics[width=\columnwidth]{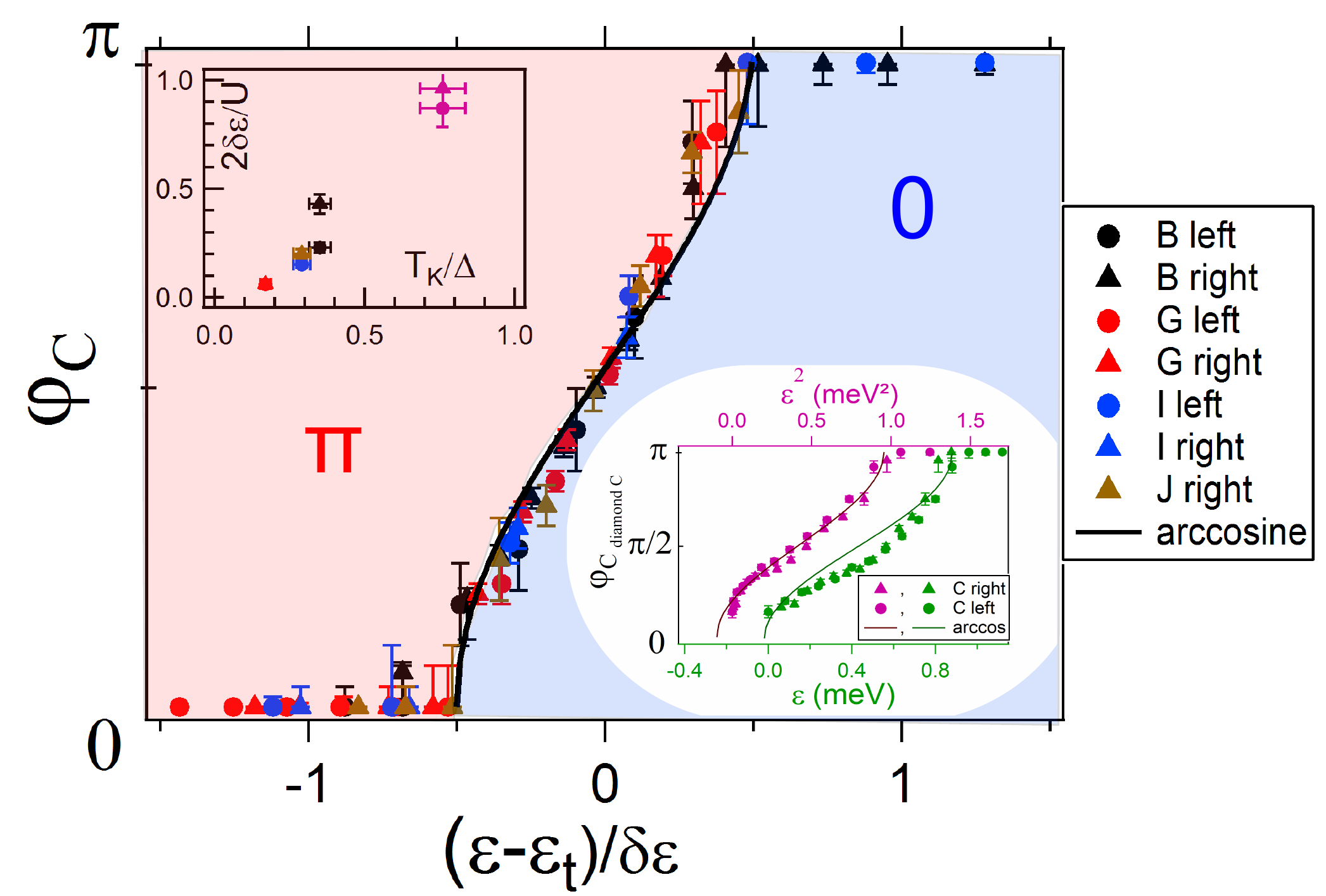}
              \end{center}
              \caption{Color. Universal scaling of the phase-induced transition.
              Critical phase $\varphi_C$ plotted as a function of $(\epsilon-\epsilon_t)/\delta\epsilon$ for the diamonds B, G, I and J (left and right sides of the diamond) such that the various curves collapse on an arccosine curve (black line). Inset left: scaling quantity $\delta\epsilon$, normalized by $U/2$, as a function of the ratio $T_K/\Delta$. It shows that $2\delta E/U$ varies linearly with $T_K/\Delta$, the quantity that controls the 0-$\pi$ transition. Inset right: same quantity for the diamond C, where the 0-$\pi$ transition is incomplete. To obtain an arccosine shape, one has to plot $\varphi_C$ as a function of $\epsilon^2$ instead of $\epsilon$ (see text).}
              \label{scaling}
              \end{figure}
 
Each transition is characterized by two parameters: the value of $\epsilon$, called $\epsilon_t$, at which the junction transits from $\pi$ to 0 for $\varphi=\pi/2$, and the width $\delta \epsilon$ of the transition. These quantities are defined on Fig. \ref{phi_c} (a) and (b) and given in table II for the concerned diamonds (B, C, G and I). $\delta \epsilon$ is found to depend strongly on the parameters of the diamonds: large transition's widths correspond to ratios $T_K(\epsilon=0)/\Delta$ close to 1 (see left inset of Fig. \ref{scaling}). 
  
 To compare these eight transitions (left and right sides of four diamonds), we plot on Fig. \ref{scaling} the critical phase $\varphi_C$ as a function of $(\epsilon -\epsilon_t)/\delta \epsilon$. For diamonds B, G and I, the scaled data fall on the same curve, with an arccosine dependence.
 This scaling can be understood within a very simple model, in which the 0-$\pi$ transition is due to a crossing of two Andreev Bound States (ABS). We make the following assumptions: (i) these ABS have a cosine shape and (ii) close to the transition, one ABS is shifted compared to the other by a quantity that is an affine function of $T_K /\Delta$.  
 
 For diamonds B, G and I, the transition occurs on the edges of the diamonds, where $T_K(\epsilon)$ can be linearized. This gives the arccosine fit on Fig. \ref{scaling}, but does not work for C (inset right of Fig \ref{scaling}), where the transition takes place close to $\epsilon=0$. Then, to the lowest order in $\epsilon$, $T_K(\epsilon)\propto \epsilon^2$, yielding $\varphi_C\propto \mathrm{arccos}(\mathrm{cste}+\epsilon^2)$ (inset of Fig. \ref{scaling} right).\newline
 
 We thus show that for all the diamonds exhibiting a complete 0-$\pi$ transition, the scaled curve $\varphi_C=f((\epsilon -\epsilon_{t})/\delta \epsilon)$ gives a robust characteristic of the phase-mediated 0-$\pi$ transition.

 \section{Effect of the two-level regime on the 0-$\pi$ transition}

\begin{figure}[h]
      \begin{center}
      \includegraphics[width=8cm]{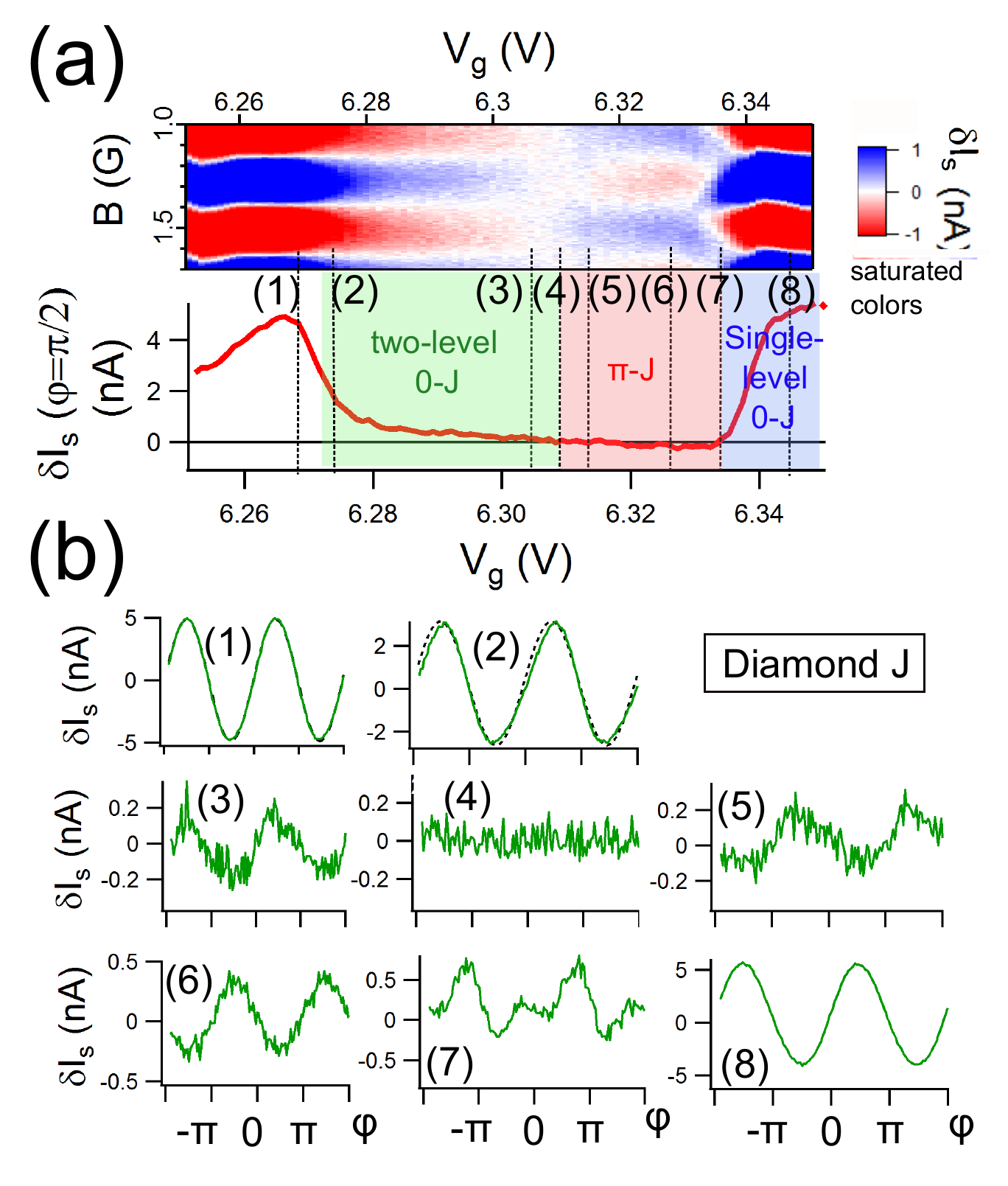}
      \end{center}
      \caption{Color. (a) Modulation of the switching current of the SQUID versus the magnetic field, proportional to the CPR, as a function of the gate voltage $V_g$, for the diamond J. The supercurrent at the transition  being very low, the color scale is saturated. 
      (b) Supercurrent versus the superconducting phase $\varphi$ at some particular gate voltage, indicated by the numbers on fig (a).Note the cancellation of the CPR (4) and the fact that the CPR (2), in the two-level 0-junction,  has a stronger anharmonicity than (8), on the degeneracy point. Dashed line on (2): guide for the eyes representing a sinus, showing that the continuous line is not perfectly harmonic.}
      \label{detail_cpr_ML}
      \end{figure}

   Now we focus on the two-level 0-$\pi$ transitions, and more particularly on diamond J, corresponding to a N=3 filling factor. The modulation of $I_s$ versus the magnetic field, proportional to the CPR, is represented on Fig. \ref{detail_cpr_ML} (a) as a function of $V_g$. CPRs are also shown for some particular values of gate voltages (Fig. \ref{detail_cpr_ML} (b)). 
   
  On the right side of the diamond, close to the N=3 to 0 degeneracy point (Fig. \ref{detail_cpr_ML}.(b) (8)), the 0-$\pi$ transition is phase-dependent, similarly to the single-level one. In addition, $\varphi_C=f((\epsilon -\epsilon_{t})/\delta \epsilon)$ collapses on the same arccosine shape as the single-level 0-$\pi$ transitions (see Fig.\ref{scaling}).
  But on the left side of the diamond, from the N=3 to 2 degeneracy point to the center of the diamond (Fig. \ref{detail_cpr_ML}.(b) (1)) , the CPR behaves as a 0-junction. The supercurrent's amplitude decreases with $V_g$ and evolves from 0 to $\pi$ continuously. Close to the transition (Fig. \ref{detail_cpr_ML}.(b) (2) to (4)), the CPR becomes slightly anharmonic. But, unlike in the single-level transition, no phase dependent 0-$\pi$ transition is observed. Note as well that the current-phase relation has a stronger anharmonicity  in the two-level 0-junction than on the charge degeneracy points, while its amplitude is smaller (Fig. \ref{detail_cpr_ML}.(b) (2), to compare with Fig. \ref{detail_cpr_ML}.(b) (1) or (8)).
  
\begin{figure}[h]
    \begin{center}
    \includegraphics[width=7cm]{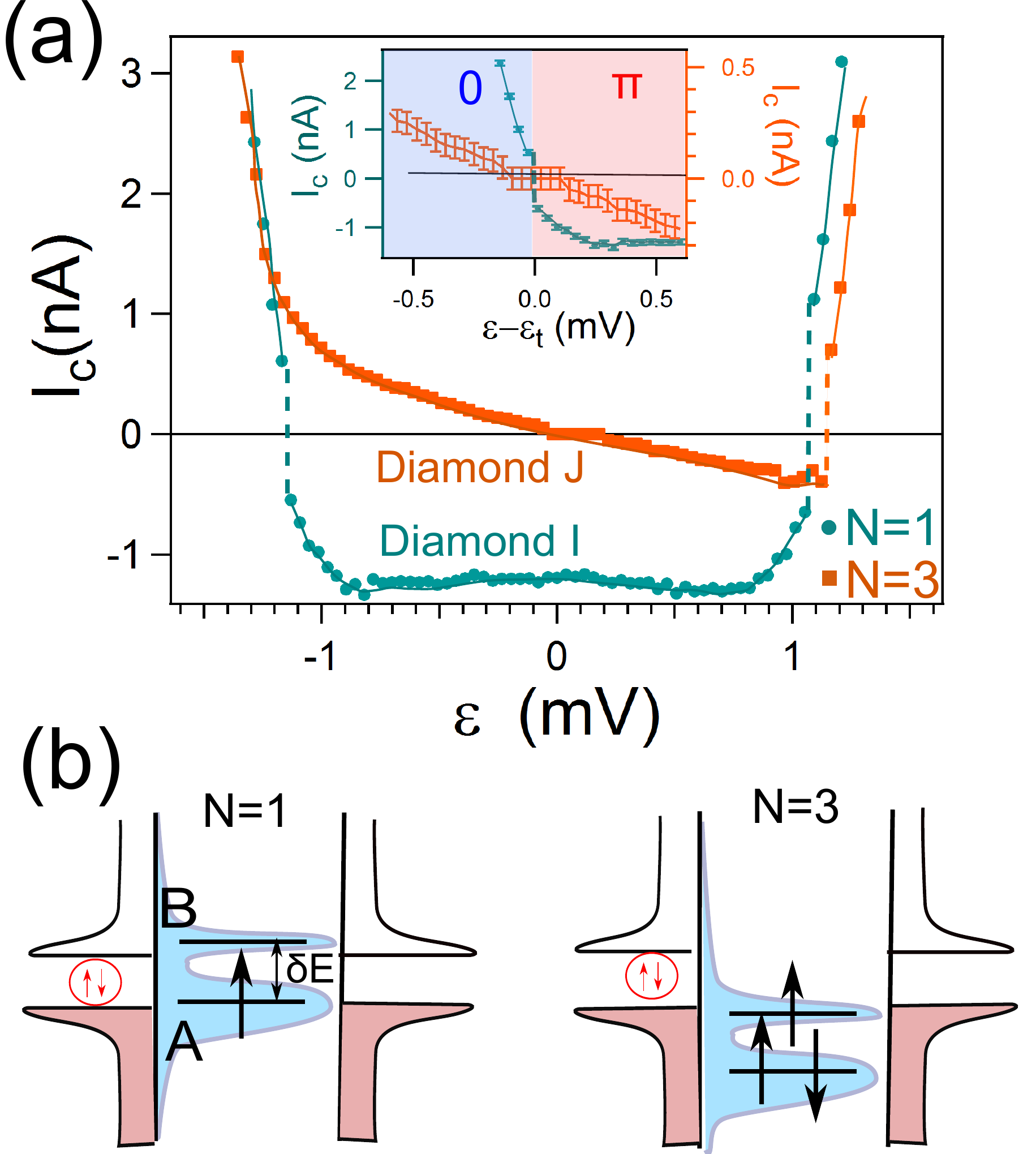}
    \end{center}
    \caption{Color. (a) Critical current $I_c$, defined as the maximum amplitude of the measured switching current, as a function of the energy level $\epsilon$. $I_c$ is defined as positive for a 0-junction, and negative for a $\pi$-junction. This quantity is plotted for the two diamonds I (N=1, blue dots) and J (N=3, orange squares), respectively in the single-level and two-level regime. The dashed lines materialize discontinuities of $I_c$, specific of first order transitions. Inset: focus on two 0-$\pi$ transitions, centered around $\epsilon_t$ (see Fig. \ref{phi_c}.(c) for the definition): at $\epsilon=-1.1\mathrm{~meV}$ in diamond I, and at $\epsilon=0.1\mathrm{~meV}$ in diamond J.
    (b) Schematic explanation of the N=1/3 symmetry breaking observed in the supercurrent. The lower energy level A is better coupled to the reservoirs than the higher one B (see text). }
    \label{SL-ML}
    \end{figure}

   To go further, we consider now the critical current $I_c$. This quantity is the maximum of the CPR and is extracted here as the maximum amplitude of the modulation of the switching current, positive for a 0-junction, and negative for a $\pi$-junction. It is represented as a function of $\epsilon$ on fig. \ref{SL-ML} (a), for diamonds I and J, respectively in the single and two-level regime. In diamond I (N=1), the phase-dependence of the single-level transitions gives rise to discontinuities of $I_c$, which characterize a first order transition \cite{Maurand2012,Vojta2006}. However, for diamond J (N=3), one of the 0-$\pi$ transition ($\epsilon=0.1\mathrm{~meV}$) does not exhibit this phase-dependence, yielding a vanishing critical current at the transition. This is not anymore a first order transition, contrary to the transition at the other side of the diamond J ($\epsilon=1.2\mathrm{~meV}$).
  
  The two other diamonds indicated by green arrows on Fig. \ref{dIdV_Is}, F and H, present the same features. This breaking of the electron/hole symmetry is observed in the three N=3 diamond where the Kondo effect is not strong enough to impose a 0-junction all-over the diamond.  \newline

This kind of gate dependence of the supercurrent is predicted in two-level quantum dots \cite{Shimizu1998} and specifically in carbon nanotubes \cite{Yu2010}, in absence of Kondo effect. The comparison of our data with Ref. \cite{Yu2010} suggests that in our experiment, the channels associated to each orbital are mixed during the transfer of Cooper pairs. This is also why we do not observe two-level induced $\pi$-junctions for even occupancies of the dot. 

To explain why this two-level behavior is not observed both for N=1 and N=3, we propose that the two orbital levels A and B (see Fig. \ref{SL-ML}.(b)) of the CNT are slightly differently coupled to the electrodes, as in Ref. \cite{Holm2008}. A detailed analysis of the gate dependence of the inelastic cotunneling peaks in the even diamond between I and J show indeed that $\Gamma_A\geqq\Gamma_B$. Following Ref. \cite{Holm2008}, we roughly evaluate $\Gamma_A-\Gamma_B\approx 0.07\mathrm{~meV}$.

 When two quasi-degenerated levels have different widths, the supercurrent is mostly carried by the broader one. We therefore expect a different physics for N=1 and N=3, as pointed out theoretically by Droste et al. \cite{Droste2012}. Here, the lower level A is more coupled to the electrodes than the higher one (B) (see Fig. \ref{SL-ML} (b)). For N=1, the unpaired electron occupies the level A and the level B is too poorly coupled to participate to the transfer of Cooper pairs: we are in a single-level situation, the junction is $\pi$. For N=3, the unpaired electron is in the poorly-coupled-level B, which thus  participates to the transport: the system is in a two-level regime. According to this interpretation, in the opposite situation of a level B better coupled than the level A, the N=1 diamond would exhibit the two-level physics instead of the N=3 diamond. 
Interestingly, while the only signature in the normal state of this breaking of the e/h symmetry is a slight change of the position of the cotunneling peaks, the supercurrent is strongly and quantitatively modified. 

 It is worth noting that this N=1/3 symmetry breaking in the supercurrent can also be seen for diamonds C-D and K-L. But the Kondo effect there is much stronger than in E-F, G-H and I-J, explaining why this unusual $\pi$-junction is not observed in D and L. Therefore, at this qualitative level, we cannot say whether they correspond to single or two-level regimes.
 
  In conclusion, our data exhibit two-level physics, qualitatively consistent with the theoretical expectations for the gate dependence. Since the orbital levels have different width, this regime is observed only when the highest occupied level is the less coupled one. We reveal as well the phase dependence of the supercurrent in this two-level regime. While single-level 0-$\pi$ transition are discontinuous first order transitions, the transition between the two-level 0 junction and the $\pi$-junction is continuous, indicating a different physical mechanism.
   
   \section{Conclusion}
   
In this article, we have measured the supercurrent of a clean carbon nanotube Josephson junction as a function of the superconducting phase. For an odd occupation of the dot, three different situations are observed. When the strength of the Kondo effect is large compared to the superconducting proximity effect, both effects cooperate and lead to a Kondo-enhanced 0-junction. When the Kondo correlations are weak, the system can sustain two kinds of $\pi$-junctions, depending on the number of levels involved in the transport (one or two). We show that when the unpaired electron is in the well coupled orbital level, the single-level regime prevails, while the two-level one occurs when the unpaired electron occupies the less coupled orbital. In other words, the electron/hole symmetry can be broken if the orbital levels are nearly degenerated and differently coupled to the electrodes. Note that the Josephson effect is far more sensitive to this symmetry breaking than the normal state conductance.

 The measurement of the current-phase relation in the single-level regime enables a detailed study of the level-crossing quantum transition driven by the superconducting phase difference. In particular, the critical phase, at which the transition happens, is shown to exhibit an universal behavior, independent of the values of the parameters of the quantum dot. In the two-level regime, this measurement shows a continuous 0-$\pi$ transition with a complete cancellation of the amplitude of the Josephson current, in contrast with the first order single-level 0-$\pi$ transition. We thus put forward the different nature of this two-level 0-$\pi$ transition.  \newline
 
 This work paves the way toward a more systematic study of the Josephson current in multi-level QD systems, for example double quantum dots or cleaner carbon nanotubes, in the presence of Kondo effect (SU(2) or SU(4)). These systems are predicted to show a very rich phase diagram \cite{Zazunov2010,Lee2010}, constituting potential new tools for superconducting circuits. 
 
 To go further, it would be very interesting to measure the current-phase relation of a CNT with a magnetic field. This would enable the study of very promising situations: 0-$\pi$ transition with a Zeeman field \cite{Yokoyama2014}, supercurrent at the singlet-triplet transition \cite{Droste2012} or Josephson junction with an arbitrary phase shift, called $\varphi_0$-junctions provided that the CNT has spin-orbit interactions and is in a multi-level regime \cite{Zazunov2009,Szombati2016}.\newline

\textit{Acknowledgments}: The authors acknowledge S. Guéron, J. Basset, P. Simon, A. Murani, F. Pistolesi, S. Florens and M. Filippone for discussions and technical help from S. Autier-Laurent. 
This work was supported by the French program ANR MASH (ANR-12-BS04-0016), DYMESYS (ANR 2011-IS04-001-01) and DIRACFORMAG (ANR-14-CE32-0003).

\section{Appendix}
\subsection{Appendix I: Bias and temperature dependence of the conductance in the normal state: determination of the quantum dot's parameters}

We give here more details about the normal state characterization of the carbon nanotube samples.

 \begin{figure}[h]
    \begin{center}
    \includegraphics[width=\columnwidth]{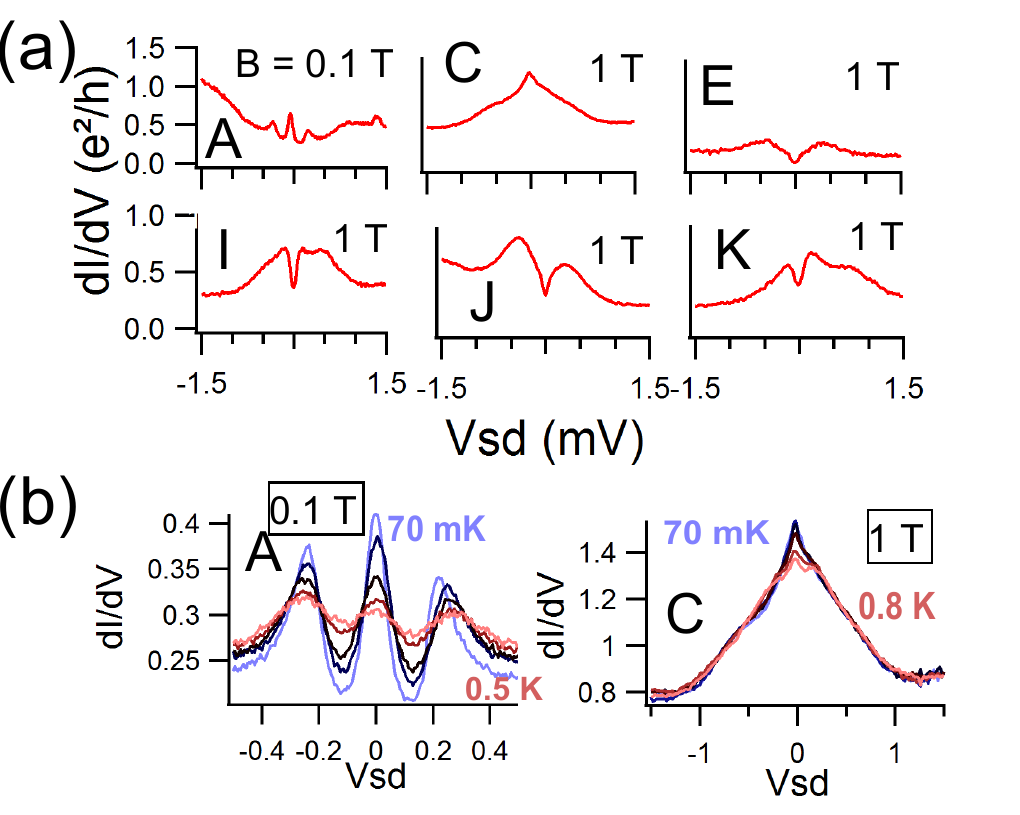}
    \end{center}
    \caption{Color. (a) Differential conductance $dI/dV_{sd}$ in the normal state as a function of $V_{sd}$ at half-filling of six different diamonds: A (S-Al) and C, E, I, J and K (S-NbAl). (b) Temperature dependence of the same quantity, for ridges A (S-Al) and C (S-NbAl). To destroy the superconductivity in the contacts, a magnetic field is applied perpendicular to the sample: $B=0.13\mathrm{~T}$ for the diamond A and $B=1\mathrm{~T}$ for the diamonds C, E, I, J and K. }
    \label{dependence}
    \end{figure} 
    
 For the sample S-Al (diamond A, Fig. \ref{dependence}), the Kondo resonance has a regular temperature dependence. In contrast, for the S-NbAl (diamonds C, E, I, J and K), the curves are more difficult to interpret for the following reasons:
 \begin{itemize}
 \item A magnetic field of 1 T is applied to destroy the superconductivity in niobium. The Kondo effect, whose $T_K$ is generally around 1K in CNT, is strongly affected by this magnetic field and the resonance can be split. This would explain the zero bias dip seen in I, J and K.
 \item These zero bias dip could also be due to some remaining superconductivity. We did not increase the field above 1T in order not to affect too much the Kondo effect.
 \item The inelastic cotunneling peaks at $V_{sd}\approx \pm 0.3-0.5\mathrm{~meV}$, present at N=1, 2 and 3 filling factor.  due to the two-level nature of the quantum dot, may be superimposed with the split Kondo peaks.
 \end{itemize}

We can nevertheless determine the parameters of the quantum dot from the stability diagram.
First, the conversion factor $\alpha$ between $V_g$ and the energy $\epsilon$ (see Fig. \ref{diamond_CNT}) is extracted for each shell from the height and width of the big diamond. Then, measuring the width of the other diamonds, we obtain U and $\Delta E$. The lift of the orbital degeneracy $\delta E$ is determined from the spacing between the inelastic cotunelling peaks in each diamond (see Fig. \ref{dependence} (a) where some of these data are shown).

For S-Al and diamond B and C of S-NbAl, the Kondo temperature is evaluated from the temperature dependence of the Kondo resonance (Fig. \ref{dependence} (b)). For the other oddly occupied diamonds of sample S-NbAl, the Kondo resonances are not clear enough to extract $T_K$ directly. It can nevertheless be determined, as in our previous work \cite{Delagrange2015}, where the conductance at zero bias of diamond I has been studied in details and compared to quantum Monte Carlo calculations taking into account the magnetic field, leading to $\Gamma=0.44\mathrm{~meV}$. For the other  diamonds, following Ref. \cite{Babic2004}, we roughly estimate the coupling $\Gamma$  from the width in $V_{sd}$ of the inelastic cotunneling peaks at N=2, and assumed that $\Gamma$ does not vary within the same shell. We estimate that these approximations are correct within an uncertainty of 20\%. 

Once $T_K$ or $\Gamma$ is evaluated, the other one is obtained from the formula \cite{Haldane1978}:
 \begin{equation}
T_K=\sqrt{\Gamma U/2}\exp(-\pi\frac{|4\epsilon^2-U^2|}{8\Gamma U})
\label{Tk}
 \end{equation}

\subsection{Appendix II: Sensitivity to the environment}
   \label{environnement_sensibility}
 
\begin{figure}[h]
       \begin{center}
       \includegraphics[width=\columnwidth]{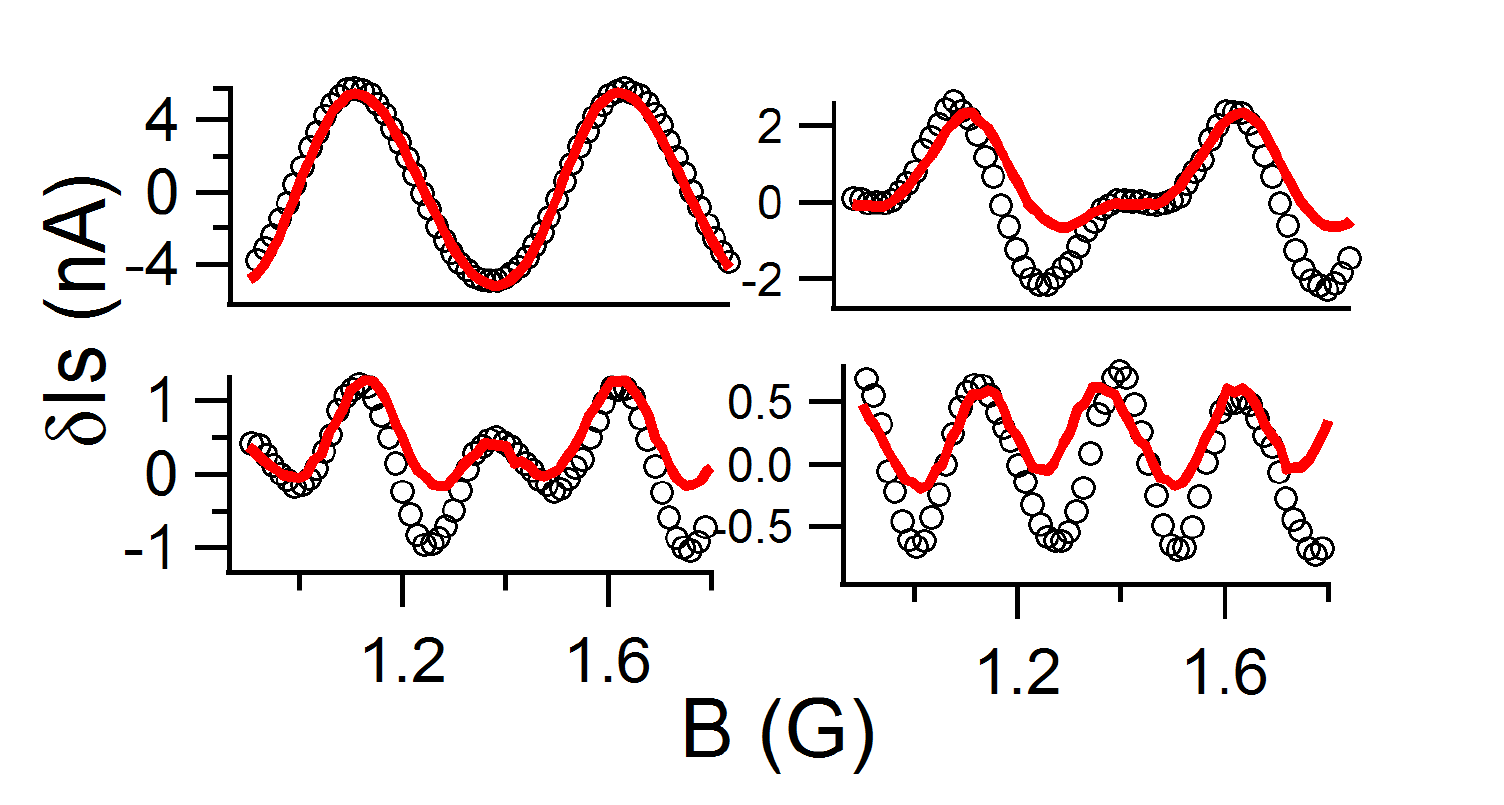}
       \end{center}
       \caption{Color. Modulation of the switching current at the 0-$\pi$ transition in diamond C with optimization of the environment (in black dots) and without (in red lines). The amplitudes of the optimized curves have been readjusted in order to compare the shapes of the curves.}
       \label{environnement}
       \end{figure}

At the 0-$\pi$ transition we noticed that, in some situations, we measured surprising CPR with an incorrect symmetry. Indeed, a CPR should respect the time-reversal symmetry that imposes the energy to be an even function of the superconducting phase, and thus the current-phase relation to be an odd function of $\varphi$. But the CPR represented in red on Fig.\ref{environnement} are obviously not odd functions of B, and thus $\varphi$, which is non-physical. One can notice that this kind of feature is also seen, but not solved, in Maurand et al. \cite{Maurand2012}. We fixed the problem optimizing the electromagnetic environment, namely fixing the polarization of a Josephson junction nearby to maximize its impedance. Note that the presence of the Josephson junction was fortuitous. We obtain the CPR's represented in black dots on Fig. \ref{environnement}, which have the good symmetry properties. 

One should note that these distortions of the CPR only happen in the 0-$\pi$ transition region, where the ABS are so closed that, if the electromagnetic environment affects the phase dynamics, phenomena such as Landau-Zener transition could also occur.

\subsection{Appendix III: Spectroscopy of Andreev Bound states.}

The CNT Josephson junction is embedded in a three arms SQUID. As detailed in a previous article \cite{Basset2014}, it allows the measurement of the differential conductance of the CNT in the superconducting state. This is done applying to the point B (see Fig. \ref{sample}) a voltage large enough to have a voltage-independent contribution of the Josephson junctions to the differential conductance.
    
  \begin{figure}[h]
           \begin{center}
           \includegraphics[width=6cm]{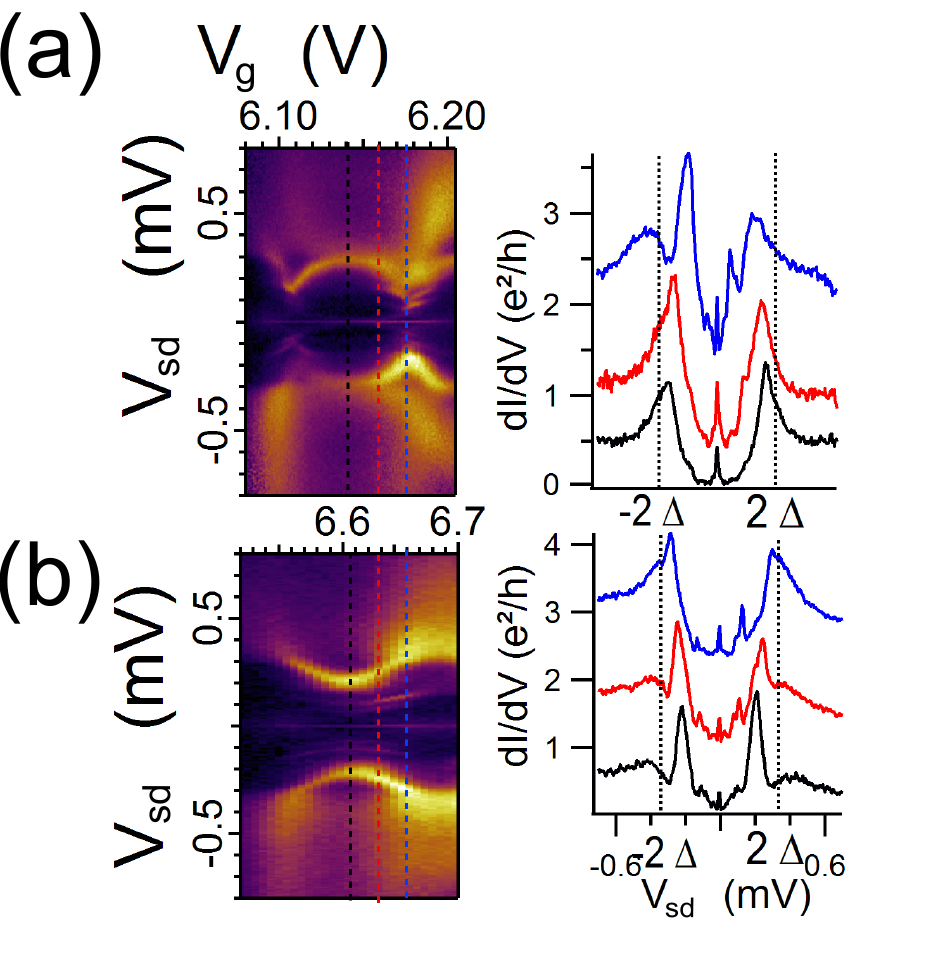}
           \end{center}
           \caption{Differential conductance as a function of the bias voltage ($dI/dV(V_{sd})$) in the superconducting state for two diamonds : (a) I ($\pi$-junction) and  (b) K (0-junction).}
           \label{dIdV_supra}
           \end{figure}
 
 In a system perfectly symmetrically coupled ($\Gamma_L=\Gamma_R$), one expects to observe multiple Andreev reflections (MAR) if the total coupling $\Gamma= \Gamma_L+\Gamma_R$ is low enough, or resonant tunneling otherwise \cite{Yeyati1997}. But if one electrode is very well coupled to the QD compared to the other ($\Gamma_L/\Gamma_R\gg 1$), one expects the poorly coupled lead to probe the density of states in the quantum dot and thus make the spectroscopy of the levels, as in Ref. \cite{Pillet2010,Chang2013}. For intermediate coupling, one observes a crossover between both \cite{Andersen2011}, as for example Ref. \cite{Eichler2007, SandJespersen2007, Deacon2010,Kim2013}.
 
 This measurement of $dI/dV(V_{sd})$ in the superconducting state (zero magnetic field) has been done for two diamonds of S-NbAl, I and K, and represented on Fig. \ref{dIdV_supra}. Diamond I is a single-level $\pi$-junction while K is a single-level 0-junction. Due to the high magnetic field, the contact asymmetry cannot be directly extracted. Thanks to QMC calculations, it has been shown in Ref. \cite{Delagrange2015} that the asymmetry of ridge I is equal to $\Gamma_L/\Gamma_R\approx 4$. We can expect a value slightly smaller for K, but not very different because they are very close in gate voltage. Because of this asymmetry, the system is not in a pure MAR regime and, according to Ref. \cite{Kim2013}, this asymmetry is large enough to interpret the conductance measurements as a spectroscopy of ABS.
 
For the two diamonds, the ABS are qualitatively different: for the ridge I, we observe a crossing of Andreev levels, characteristic of $\pi$-junctions as explained on Fig. \ref{sample}. For the diamond K, because there is no $\pi$-junction, the Andreev-levels do not cross.
Even though this kind of measurement has already been done in a similar system in Ref. \cite{Kim2013}), where this spectroscopy of Andreev levels had been done on the same sample as the critical current measurement, here the spectroscopy is done in parallel with the current-phase relation measurement.

 It should be noted that this measurement of the conductance as a function of the bias voltage cannot give any information about the phase-dependent 0-$\pi$ transition. Indeed, as soon as the JJ is voltage biased, the superconducting phase is not controlled anymore and varies as $d\varphi/dt=2eV/\hbar$ \cite{Tinkham1996}. 
 
 \subsection{Appendix IV: SU(4) Kondo effect?}
    
\begin{figure}[h]
            \begin{center}
            \includegraphics[width=8cm]{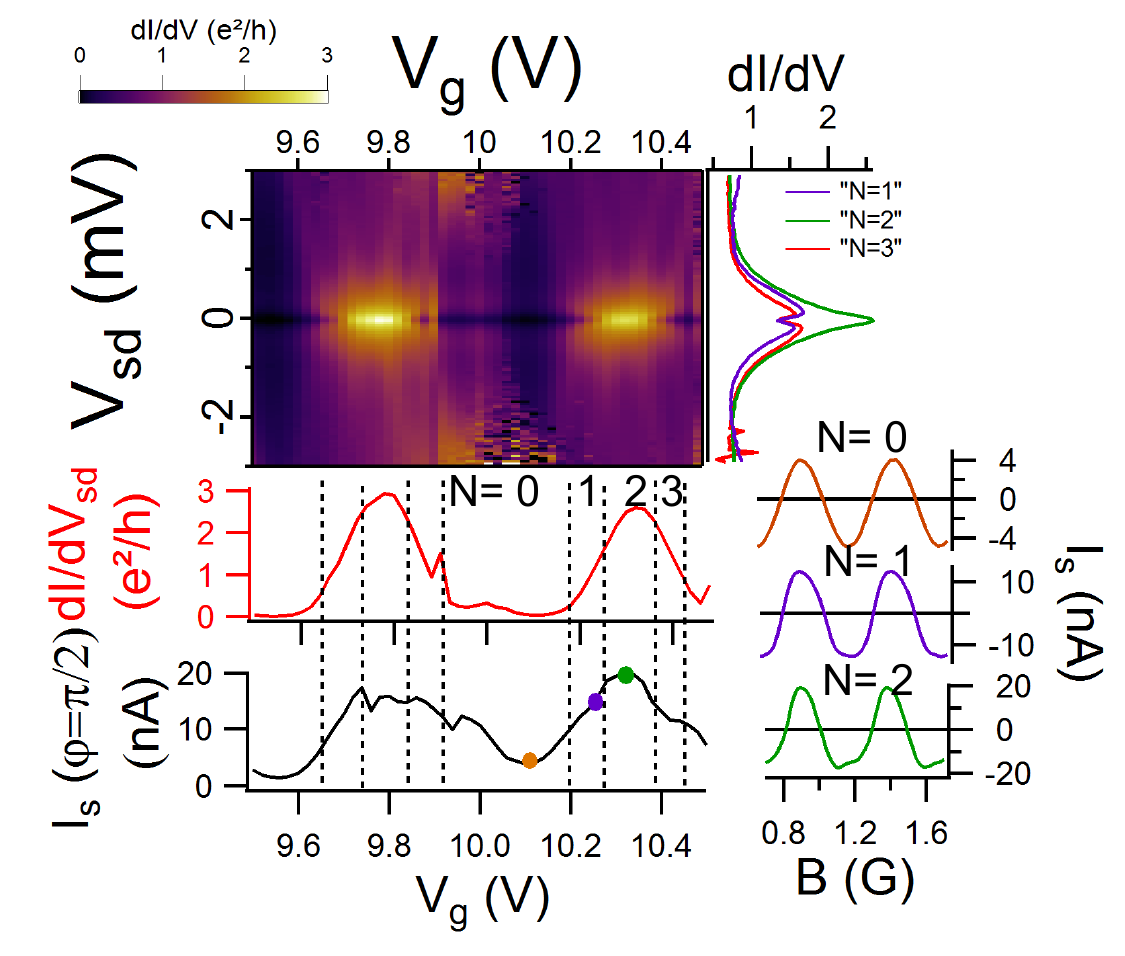}
            \end{center}
            \caption{Color. Data in the normal and superconducting state for a probable SU(4) Kondo effect zone. Colorplot: differential conductance as a function of $V_{sd}$ and $V_g$. A magnetic field of 1T is applied perpendicularly to the sample to destroy the superconductivity in the contacts. On the right are plotted vertical cuts of the color-plot, yielding $dIdV(V_{sd})$ at the middle of N=1, N=2 and N=3 diamonds.  Red curve: $dIdV(V_g,V_{sd}=0)$ in the normal state. Black curve: supercurrent $\delta I_s(V_g)$ for a superconducting phase difference of $\varphi=\pi/2$. On the right are plotted current-phase relations at the center of N=0, 1 and 3 diamonds. }
            \label{su4}
            \end{figure}

We present here measurements on sample S-NbAl in a range of gate voltage where the coupling between the carbon nanotube to the electrodes is larger. On the stability diagram represented on Fig. \ref{su4}, we can guess diamond-like features, showing a succession of three small diamonds, one big, and again three small ones. However, contrary to Fig. \ref{dIdV_Is}, the non-zero conductance at zero bias, \textit{i.e}. the Kondo effect, spreads all over the three small diamonds (N=1, 2 and 3). In addition, the zero bias conductance in the N=2 diamond reaches $2.5e^2/h$.

These features seem to indicate SU(4) Kondo effect, where two independent degrees of freedom, here the spin and the orbital pseudo-spin, are involved \cite{Choi2005,Makarovski2007}. We were not able to perform the magnetic field study of the conductance, that would confirm the presence of this effect.
     
Still, we measure as well the current-phase relation in this regime (Fig. \ref{su4}). First, the junction is in a 0-state for all gate voltages. We notice as well that the supercurrent is very high, reaching 20 nA in the N=2 diamond around $V_g=10.35 V$. Such an increase of the critical current compared to the Kondo SU(2) case is expected since, in Kondo SU(4) at N=2 filling, two perfectly transmitted channels are open (instead of one in Kondo SU(2)). 
Unfortunately, due to this high value of the supercurrent, in this regime our experimental setup cannot be considered anymore as an asymmetric SQUID. The CPR measurement is therefore not reliable, which may explain the unusual anharmonicities observed at N=2.

\end{document}